\documentclass{article}

\usepackage{epsf}

\def\abstract#1{\vskip 7mm 
        \begin{center}{\large Abstract}\par \smallskip
                \begin{minipage}[c]{12cm}
                        \small #1
                \end{minipage}
        \end{center}
}
\def\title#1{\begin{center}{\Large\bf #1}\end{center}}
\def\author#1{\vskip 5mm \begin{center}{#1}\end{center}}
\def\address#1{\begin{center}{\it #1}\end{center}}
\makeatletter
\@ifundefined{lesssim}{}{}
\@ifundefined{gtrsim}{\def\gtrsim{\mathrel{\mathpalette\vereq>}}}{}
\def\vereq#1#2{\lower3pt\vbox{\baselineskip1.5pt \lineskip1.5pt
\ialign{$\m@th#1\hfill##\hfil$\crcr#2\crcr\sim\crcr}}}
\makeatother

\begin{document}

\title{%
  Relativistic SPH Calculations of Compact Binary Mergers
}
\author{%
  Frederic Rasio\footnote{Email: rasio@northwestern.edu},
  Joshua Faber\footnote{Email: jfaber@uiuc.edu},
  Shiho Kobayashi\footnote{Email: shiho@gravity.psu.edu},
  and Pablo Laguna\footnote{Email: pablo@astro.psu.edu }
}
\address{%
  $^1$Department of Physics and Astronomy, Northwestern University, Evanston, IL 60208, USA\\
  $^2$Department of Physics, University of Illinois, Urbana, IL 61801, USA\\
  $^3$Center for Gravitational Wave Physics,
     Pennsylvania State University, University Park, PA 16802, USA\\
  $^4$Department of Astronomy and Astrophysics and Department of
      Physics, Pennsylvania State University, University
		 Park, PA 16802, USA
}

\abstract{
We review recent progress in understanding the
hydrodynamics of compact binary mergers using relativistic
smoothed particle hydrodynamics (SPH) codes. Recent results
are discussed for both double neutron stars and black hole -- neutron star binaries. The gravitational wave signals from these mergers
contain a lot of information about the masses and radii of the stars 
and the equation of state of
the neutron star fluid. For black hole -- neutron star mergers,
the final fate of the fluid and the gravitational wave emission depend
strongly on the mass ratio and the black hole spin. }

\section{Introduction}

Coalescing compact binaries containing neutron star (NS) or
black hole (BH) components
are the leading candidates to become the first observed sources of
gravitational waves (GW). With LIGO, GEO, and TAMA all taking scientific
data, and VIRGO in the commissioning stage, it is now
more important than ever to develop quantitatively accurate 
predictions of the GW signals we expect to observe from the 
binary merger process.  These GW signals contain crucial information about 
the masses, NS radii, and NS equation of state (EOS).

Calculations of binary NS coalescence have been performed for many
years, beginning almost 20 years ago with studies in Newtonian 
gravity (see, e.g.,
Nakamura 1994; Rasio \& Shapiro 1999; for reviews of early work).  It was
recognized all along, however, that general relativity (GR) must play
an important role, given the strong
gravitational fields and relativistic velocities involved,
particularly near the end of the merging process.
As a result, increasingly sophisticated gravitational formalisms have
been used in combination with 3-D hydrodynamic calculations, starting with  
post-Newtonian (PN) treatments (Oohara \& Nakamura 1992; Shibata 1997;
Lombardi, Rasio, \& Shapiro 1997; Faber \& Rasio 2000, 2002; Ayal et al.\ 2001). Several recent studies were based on the
PN hydrodynamics formalism developed by Blanchet, Damour \&
Sch\"afer (1990),
which includes all lowest-order 1PN effects as well as the lowest-order
dissipative effects from gravitational radiation reaction. 
Most recently, calculations have been performed in full GR, using
the techniques of 3-D numerical relativity (Shibata \& Uryu 2000, 2001;
Shibata, Taniguchi, \& Uryu 2003; Miller, Gressman, \& Suen 2004).  
These are plagued by numerical
difficulties, but allow one to address fundamentally new and
important questions, such as the
stability of the final merger remnant to gravitational collapse.

Unfortunately, while considerably easier to implement numerically,
 the PN approximation breaks down
during the merger when higher-order relativistic effects grow
significant.  A middle ground is provided by the
conformally flat (CF) approximation to GR, developed originally by 
Wilson and collaborators (see, e.g., Wilson, Mathews \& Marronetti 1996).
This approximation is used in essentially all quasi-equilibrium 
calculations of initial data for compact binary mergers (NS--NS, BH--NS,
and BH--BH; see, e.g., Faber et al.\ 2002;
Grandcl{\'e}ment, Gourgoulhon, \& Bonazzola 2002).
It has also been used more recently for full 
hydrodynamic merger calculations
(Oechslin, Rosswog, \& Thieleman 2002; Faber, Grandcl{\'e}ment, \& Rasio 2004).
An overview of the latest results from Faber et al.\ (2004) is given in Sec.~3.
The CF approximation
includes much of the non-linearities inherent in GR, but reduces the Einstein
field equations to a
set of coupled, non-linear, elliptic equations, which can be
solved numerically using well-understood, robust techniques.  
While this approach cannot reproduce the exact GR solution for a
general 3-D matter configuration, 
it is exact for spherically symmetric systems, and, in practice, 
yields solutions
that often agree closely with those calculated using
full GR (Mathews, Marronetti, \& Wilson 1998; Shibata \& Sekiguchi 2004).
Most remarkably,
CF solutions for BH--BH binaries in quasi-equilibrium agree
extremely well with those obtained from the most accurate, high-order
PN treatment (Damour, Gourgoulhon, \& Grandcl{\'e}ment 2002).

In contrast to NS--NS merger calculations, relatively little work has
been done on BH--NS mergers. Newtonian hydrodynamic calculations
for a polytrope around a point mass (e.g., Lee 2001) are of limited 
applicability to
the highly relativistic BH--NS systems. Relativistic
quasi-equilibrium calculations of BH--NS binaries near the Roche limit
have been performed recently by several groups
(Baumgarte, Skoge, \& Shapiro 2004; Ishii, Shibata, \& Mino 2005).
Some initial results from our own relativistic hydrodynamic
study of BH--NS mergers are presented in Sec.~4.
In spite of the great astrophysical importance of BH--NS binaries, our 
theoretical understanding of these systems lags behind that of 
NS--NS binaries by at least 10 years. One key question concerns the
timescale of the tidal disruption and whether quasi-stable mass transfer
from a NS to a BH can ever occur. This has
important consequences for the detectability of the GW signal associated
with the tidal disruption. This signal carries information about the NS EOS,
and could be easier to detect with broadband interferometers
than NS--NS merger signals (Vallisneri 2000).

A systematic study of BH--NS binary mergers 
will require exploration of a parameter space much larger than for NS--NS binaries. In particular, a wide range of 
mass ratios should be considered, $q=M_{\rm h}/M_{\rm ns}\simeq 1-20$, since BH masses are 
not expected to be concentrated in a narrow range (observations of
BH X-ray binaries suggest $M_{\rm h}\simeq 5-15\,M_\odot$, but this may be
biased by observational selection effects).
Newtonian results suggest that for very low-mass BH ($q\simeq1$) the NS is
quickly disrupted after the system reaches the ISCO, while for
more massive BH ($q\gtrsim 5$) the behavior is more erratic and complete tidal 
disruption of the NS may only occur after a large number of orbital periods 
and repeated brief episodes of mass transfer (Davies, Levan, \& King 2005). 
In most cases, roughly one third of the NS mass is left in a thick torus around
the BH while the rest is quickly accreted. While these results may be (at best)
qualitatively correct, it is clear that relativistic effects must play an
important role in the evolution. For a massive BH in particular, the tidal
disruption of the NS takes place entirely in a strong field region
(see Sec.~4). 

Although this paper will focus on the GW aspects of the binary coalescence
problem, the astrophysical motivation for the study of this problem is
much broader. The ejection of nuclear matter during mergers
has been discussed as a possibly major source of heavy ``r-process''
elements in galaxies (Lattimer \& Schramm 1974; Rosswog, Freiburghaus,
\& Thielemann 2001).
Many theoretical models of gamma-ray bursts (GRBs) 
have also relied on coalescing compact 
binaries to provide the energy of GRBs at
cosmological distances (e.g., Eichler et al.\ 1989; Narayan, 
Paczy\'nski, \& Piran 1992; M\'esz\'aros \& Rees 1992). 
Currently these models all assume that the coalescence leads
to the formation of a rapidly rotating Kerr BH surrounded by a torus
of debris (e.g., van Putten \& Levinson 2002). 
Energy can then be extracted either from the rotation of the BH or from
the material in the torus so that, with sufficient beaming, the
gamma-ray fluxes observed from even the most distant GRBs could be
explained. Here also, it is important to
understand the hydrodynamic processes taking place during the final 
coalescence before making assumptions about its outcome. For example,
contrary to widespread belief, it is not clear that the coalescence of
two $1.4\,M_\odot$ NS will form an object that must collapse to a BH
(Sec.~3).

\section{Smoothed Particle Hydrodynamics}

The hydrodynamic calculations of compact binary coalescence
described below in Sec.~3 and~4 were done using two different relativistic 
smoothed particle hydrodynamics (SPH) codes, which we will refer to
as the ``Northwestern code'' and the "Penn State code.'' SPH is a Lagrangian,
particle-based method ideally suited to calculations involving
self-gravitating fluids moving freely in 3D. 

Because of its Lagrangian nature,
SPH presents some clear advantages over more traditional 
grid-based methods for calculations of stellar interactions. Most
importantly, fluid advection, even for stars with a sharply defined surface
such as NSs, is accomplished without difficulty in SPH, since the
particles simply follow their trajectories in the flow. In contrast, 
to track accurately the orbital motion of two stars across a large 3-D grid
can be very difficult, and sharp stellar surfaces often require a special
treatment (to avoid ``bleeding''). SPH is
also very computationally efficient, since it concentrates the
numerical elements (particles) where the fluid is at all times,
not wasting any resources on empty regions of space. For
this reason, with given computational resources, SPH can provide higher
averaged spatial resolution than grid-based calculations, although
Godunov-type schemes such as PPM typically provide better 
resolution of shock fronts. This is certainly not a decisive advantage for
binary coalescence calculations, where no strong shocks ever develop.
SPH also makes it easy to track the hydrodynamic ejection of matter to 
large distances from the central dense regions. Sophisticated nested-grid 
algorithms are necessary to accomplish the same with grid-based
methods. 

The first relativistic version 
of the Northwestern code (using PN gravity) was developed by 
Faber \& Rasio (2000, 2002). For a detailed
description of the purely hydrodynamic aspects of this code, including the 
results of many test calculations, see Lombardi et al.\ (1999).
In addition to binary coalescence problems this code has also been 
used extensively in Newtonian gravity
to study many problems related to stellar collisions in dense star clusters
(e.g., Sills et al.\ 2001).
Our current relativistic version of the code is based on the CF approximation to 
full GR (Faber et al.\ 2004). We solve the elliptic field equations using 
spectral methods. Specifically, our SPH code was coupled to the
software developed by the Meudon group, which is publicly 
available as part of the
C++ LORENE library ({\tt www.lorene.obspm.fr}). 
In contrast to finite-difference schemes, the fields are 
represented in terms of expansions over analytic basis functions. The 
effects of all operators, such as derivatives, are then represented 
by matrix multiplication on the coefficients of the expansion. In LORENE, 
the tensorial basis of decomposition is Cartesian, but 
each component is given in terms of spherical coordinates. 
In the current version, 
the usual basis of decomposition contains trigonometric polynomials 
or spherical harmonics for the dependence on the angles and 
Chebyshev polynomials for the dependence on the radial coordinate
(Bonazzola et al.\ 1999).

The original version of the Penn State code is described by
Laguna, Miller, \& Zurek (1993). This code models relativistic 
fluid flow in a {\it fixed\/} 
curved spacetime geometry assuming that (a) the background spacetime is static
(e.g., Schwarzschild or Kerr BH), and (b) the density of mass-energy of
the fluid is sufficiently small that a Newtonian model of the fluid's
self-gravity is an adequate approximation. The code was calibrated with several
1-D benchmarks (Laguna et al.\ 1993) such as relativistic shock
tubes, dust infall onto a BH, and Bondi collapse. It has been used recently
to study the tidal disruption of main sequence stars by a
supermassive BH (Laguna et al.\ 1993b; Kobayashi et al.\ 2004). 
For our BH--NS merger calculations we use the Kerr-Schild form of the 
Kerr (or Schwarzschild) metric to
describe a fixed background. The coordinates smoothly penetrate the
horizon of the holes, and they are suitable for studying 
the evolution of stars orbiting closely around the horizon or even
falling into it. An absorbing boundary is set up well inside the horizon,
and SPH particles crossing this boundary are removed from the
calculation before they hit the BH singularity. This ensures a long-term,
stable numerical evolution of the system. For obvious reasons this code
should apply most accurately to BH--NS binaries in which the BH mass is much
larger than the NS mass.

\section{Neutron Star Binaries}

%Fig 1
\begin{figure}[tbp]
\epsfxsize=16cm
\centerline{\epsfbox{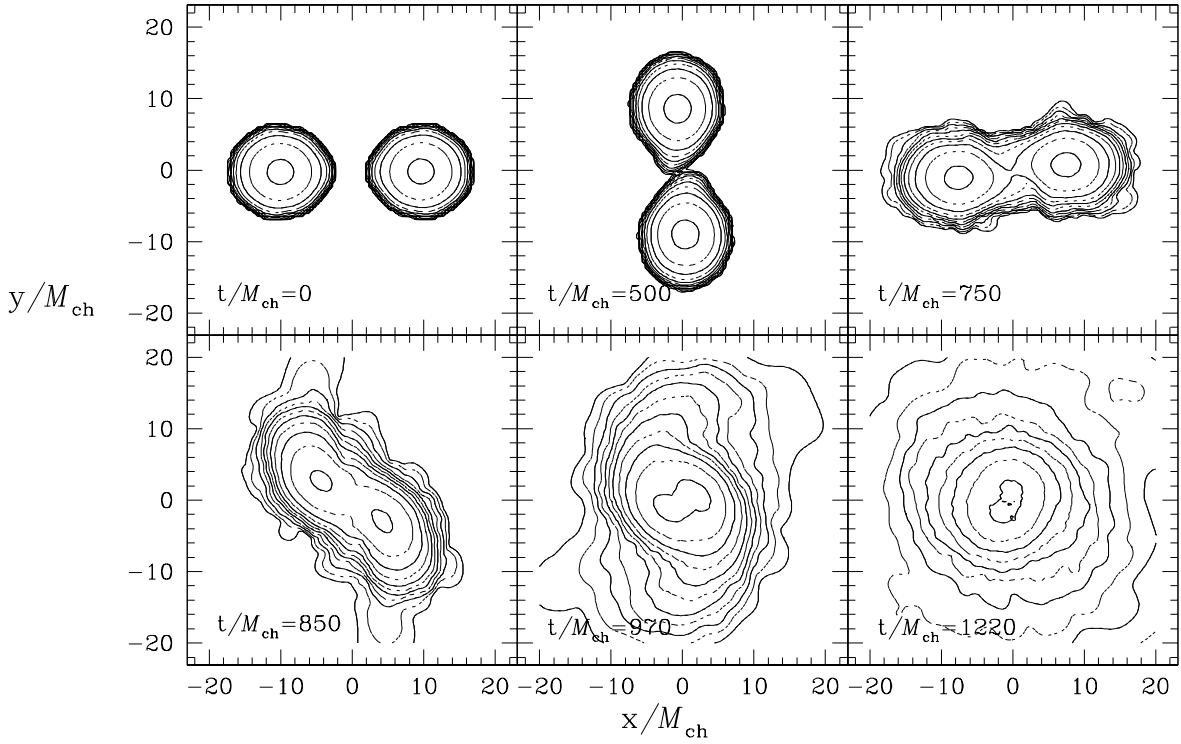}}
\caption{Final merger of a NS--NS binary. Units are defined such that
  $G=c=1$, and $M_{\rm ch}$ is the chirp mass of the system.
  Density contours are shown in the equatorial ($x-y$) plane of the
  system, logarithmically spaced with two contours per
  decade. We first see the development of significant tidal lag, followed by 
  an ``off-center collision'' that leads to the formation of a vortex sheet
  at the interface between the two stars
  and a small amount of matter ejection. The final configuration is a
very rapidly and differentially rotating, spheroidal NS, which is stable
  against gravitational collapse. This calculation was performed using $10^5$
  SPH particles in the CF approximation of GR. The initial condition 
  consists of two identical, non-spinning (irrotational) NSs with a $\Gamma=2$
  polytropic EOS, in a quasi-equilibrium circular orbit outside the
  ISCO. The orbital rotation is in the
  counter-clockwise direction. } 
\end{figure}

A typical NS--NS binary merger, calculated with the Northwestern
SPH code in CF gravity, is illustrated in
Fig.~1 (from Faber et al.\ 2004).  
The initial phase of the evolution corresponds to the end of the binary inspiral,
but with finite-size effects becoming important. In particular,
a significant ``tidal lag'' (misalignment of the stellar long axes
with respect to the binary axis) develops before contact. When
physical merging begins, the two stars come together in an off-axis manner, with a
very small amount of mass running along the surface interface before
being spun off the newly forming remnant.  This small amount of
ejected matter, representing less than $1\%$ of the total system mass
in this particular calculation,
remains gravitationally bound, forming a tenuous halo around the
rapidly and differentially rotating remnant, which is stable to
gravitational collapse.

We calculate the GW signal produced during the merger in the
lowest-order quadrupole limit, finding good agreement with
other recent relativistic calculations (Faber et al.\ 2004). 
Prior to final merger, we observe a ``chirp'' signal, as the frequency and
amplitude both increase while the two NSs approach each other.  After the
merger, there is a period of strong, modulated high-frequency
emission, which damps away as the remnant relaxes toward a spheroidal
shape.  

% Fig 2
\begin{figure}[tbp]
\epsfxsize=14.5cm
\centerline{\epsfbox{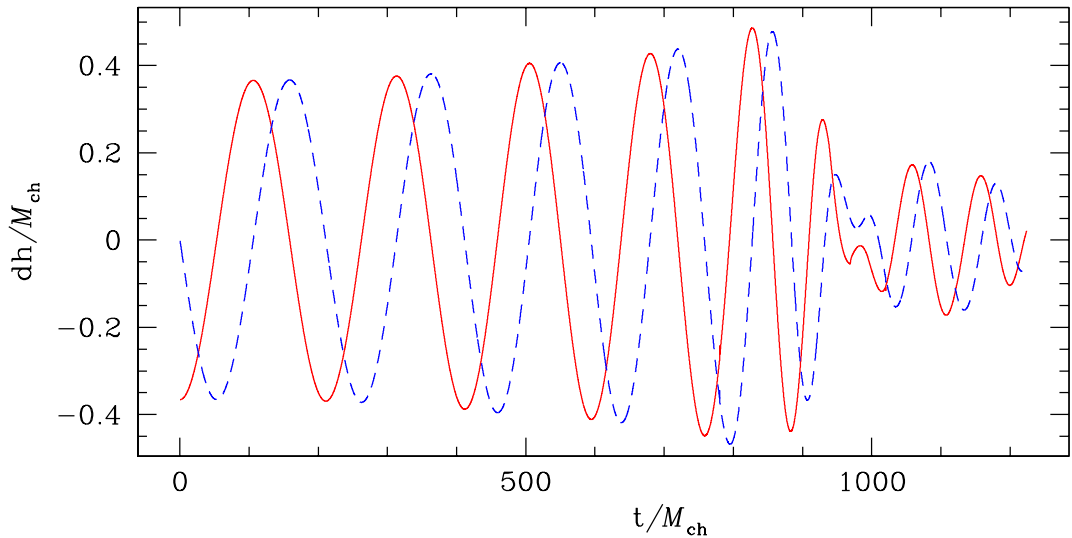}}
\caption{Gravitational wave signal in the $h_+$ (solid line) and
  $h_\times$ (dashed line) polarizations, for an observer located a
  distance $d$ from the system along the rotation axis.  Units are as
  in Fig.~1.  We see the end of the inspiral chirp signal, followed 
  by a lower-amplitude, modulated burst of high-frequency emission
  during the final merger.}
\end{figure}

While the time-dependence of the GW signal is important, it is
more useful in practice to look at the frequency dependence of the
signal, and, in particular, the energy spectrum $dE_{\rm GW}/df$.  
Indeed, one can show (Faber et al.\ 2002) that the changes in
the total energy along quasi-equilibrium binary configurations for NS
models with different compactness values $M/R$  should
leave an imprint in the energy spectrum in the form of a ``break,''
where the spectrum plunges rapidly below the power-law extrapolation of the
low-frequency, slow inspiral spectrum ($dE_{\rm GW}/df\propto f^{-1/3}$).
This corresponds to the sudden acceleration of the orbital decay
as the NS binary approaches the point of onset of dynamical instability
along a quasi-equilibrium sequence (Lai, Rasio, \& Shapiro 1994;
Lombardi et al.\ 1997).
This is potentially detectable by 
narrow-band detectors in advanced interferometers. Using a
parameterized model of these ``break frequencies,''
Hughes (2002)
showed that the NS radius could be determined to within a few
percent with at most $\sim 50$ Advanced LIGO observations of NS mergers,
and perhaps far fewer for optimal parameter values.

Since our calculation was started from a proper quasi-equilibrium initial
condition, we
can easily attach the inspiral waveform onto our calculated merger waveform at
the point where the binary reaches the correct infall velocity.
We note, however, that the exact point where the crossover is made
has virtually no effect on the resulting spectrum.
In Fig.~3, we show (as the solid line)
a complete and {\it self-consistent\/} relativistic spectrum for
a binary NS merger.  At low frequencies, the primary contribution is 
from the inspiral signal, and at high frequencies, from the calculated merger
waveform.  The short-dashed line shows the Newtonian
power-law result for two point masses, and the long-dashed curve shows 
the fit we
find from our quasi-equilibrium sequence data (following Faber et al.\ 2002).  
 We see excellent agreement up until the peak at frequencies around
 $M_{\rm ch}f_{\rm GW}\simeq 0.007-0.009$.  This peak represents the
``piling up'' of energy corresponding to the moment when the stars 
make contact and the radial infall rate
slows down in response to pressure building up.  The second peak, at 
$M_{\rm ch}f_{\rm GW}\simeq 0.010-0.011$, represents emission from the
ringdown of the merger remnant.  It is likely that we underestimate
the true height of this second peak somewhat, since we assume that the GW
signal after our calculation ends continues to damp away exponentially, 
but the detection of this high-frequency peak will be quite challenging
anyway.  

% Fig 3
\begin{figure}[tbp]
\epsfxsize=17cm
\centerline{\epsfbox{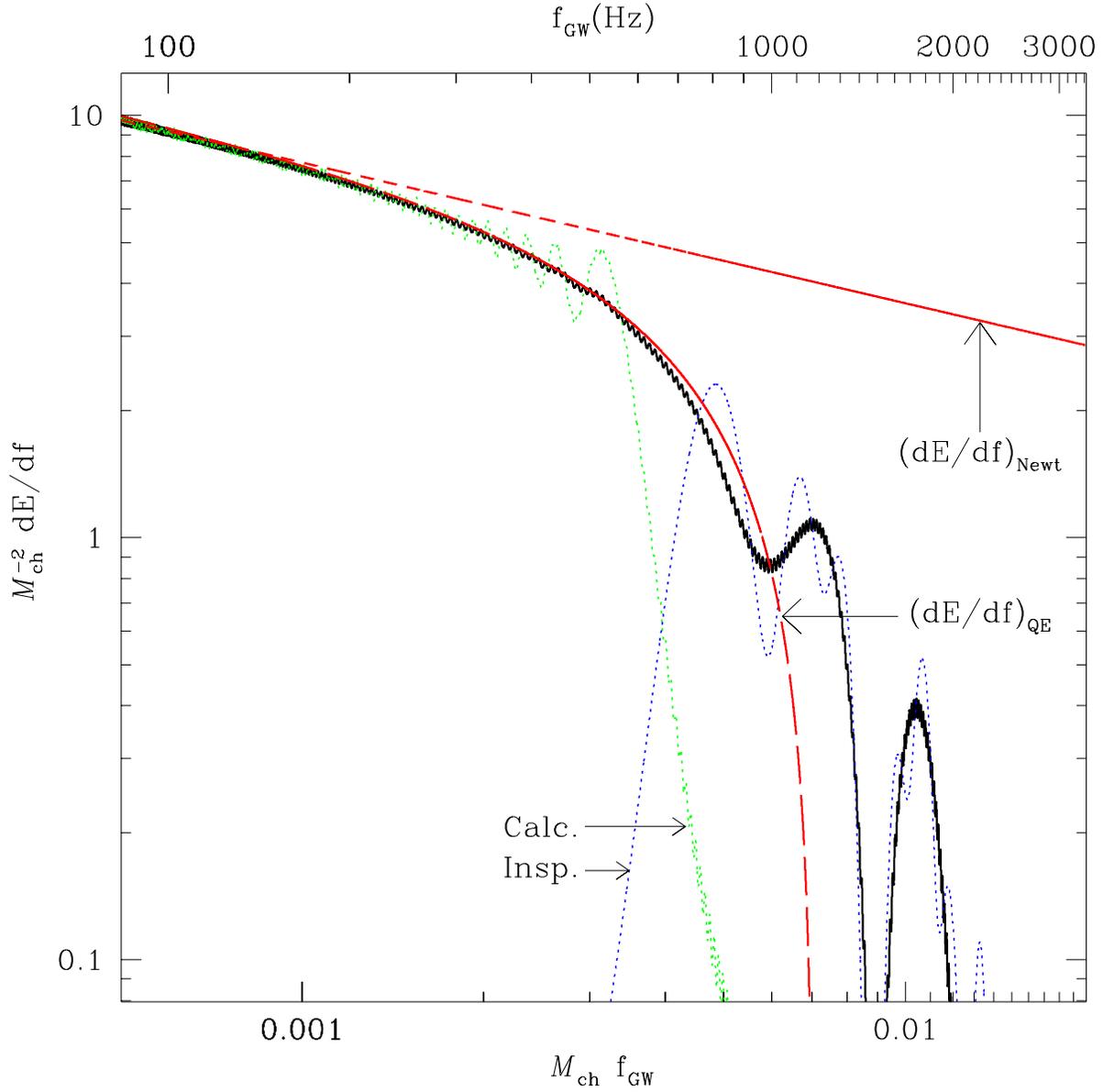}}
\caption{Gravitational wave energy spectrum, $dE/df$, as a
  function of GW frequency, $f_{\rm GW}$.
  Units are defined such that
  $G=c=1$, and $M_{\rm ch}$ is the chirp mass of the system.
  The dotted lines at high and low
  frequencies show the components contributed by our calculated signal
  and the quasi-equilibrium inspiral component, respectively.  Also shown 
  are the Newtonian
  point-mass energy spectrum (short-dashed line), 
  and the semi-analytic result derived from a quasi-equilibrium 
  binary NS sequence (long dashed line; Faber et al.\ 2002).  On the
  upper horizontal axis, we also show the corresponding frequencies in 
  Hz assuming that each
  NS has a mass of $1.4\,M_\odot$.  We see that the ``break frequency''
  occurs well below $\sim 1\,$kHz, where ground-based interferometers
  retain good sensitivity (above $\sim 1\,$kHz laser shot noise leads
  to a rapid loss of sensitivity).}
\end{figure}

The final fate of a NS--NS merger depends crucially on the NS EOS,
and on the extraction of angular momentum from the system during the 
final merger. For a stiff NS EOS, it is by no means
certain that the core of the final merged configuration would collapse
on a dynamical timescale to form a BH. One reason is that the Kerr
parameter $a/M$ of the core may exceed unity for an extremely stiff
EOS (Baumgarte et al.\ 1998), although recent 
hydrodynamic calculations suggest that this is never the case
(see, e.g., Faber \& Rasio 2000). 
More importantly, the rapidly rotating core may in fact be 
dynamically stable. Take the obvious example of a system containing two 
identical $1.4\,M_\odot$ NSs. The total baryonic mass of the system
for a stiff NS EOS is then about $3\,M_\odot$. 
Most modern (stiff) nuclear EOS allow stable,
maximally rotating NS with baryonic masses exceeding
$3\,M_\odot$ (Cook, Shapiro, \& Teukolsky 1994), i.e., above the mass
of the final merger core, even if we entirely neglect fluid ejection. 
{\it Differential\/} rotation can further increase 
this maximum stable 
mass very significantly (Baumgarte, Shapiro, \& Shibata 2000).
Using realistic nuclear EOS,  Morrison, Baumgarte, \& Shapiro (2004)
show that the merger remnants of all NS binaries observed in our Galaxy as
binary pulsars could be supported stably against prompt collapse by
rapid differential rotation. Recent full GR calculations of NS
binary mergers in 3-D numerical relativity have also suggested that
at least some merger remnants with stiff EOS are stable to collapse 
(Shibata, Taniguchi, \& Uryu (2003).

However, for slowly rotating stars, the same EOS give
maximum stable baryonic masses in the range $2.5-3\,M_\odot$.
Thus the final fate of the merger depends critically on its rotational
profile and total angular momentum. Many processes such as
electromagnetic radiation, neutrino emission, or the development of
various secular instabilities (e.g., r-modes), which may also lead to angular
momentum losses, take place on timescales much longer than the dynamical
timescale (see, e.g., Baumgarte \& Shapiro 1998, who show that
neutrino emission is probably negligible). These processes are
therefore decoupled from the hydrodynamics of the merger.
Unfortunately their
study is plagued by many fundamental uncertainties in the microphysics.
Duez et al.\ (2004) have demonstrated for the first time in full GR 
how viscosity and magnetic fields can drive 
a differentially rotating, ``hypermassive'' NS merger remnant toward uniform 
rotation, possibly leading to delayed collapse to a BH. This delayed
collapse would be accompanied by a secondary burst of GW. 

The question of the final fate of the merger also depends crucially
on the evolution of the fluid vorticity during the coalescence.
Close NS binaries  are likely to be {\it nonsynchronized\/}. Indeed,
the tidal synchronization time 
is almost certainly much longer than the orbital decay
time (Kochanek 1992; Bildsten \& Cutler 1992).
For NS binaries that are far from synchronized,
the final coalescence involves complex hydrodynamic 
processes (Rasio \& Shapiro 1999).
In the simple case of an irrotational system (containing stars that are
non-spinning at large separation),
the two sides appear to be counter-spinning in the corotating
frame of the binary. At the moment of first contact, a {\it vortex sheet\/}  
is formed. Such a vortex sheet is Kelvin-Helmholtz unstable on all 
wavelengths and the hydrodynamics is therefore very challenging
to model accurately given the limited spatial
resolution of 3-D calculations (see Faber \& Rasio 2002 for a high-resolution
SPH calculation using $10^6$ particles and focusing on this aspect of
the problem).

The final fate of the merger could depend crucially on the evolution
of this shear flow inside the merging stars. In particular, if the 
shear persists and can support a highly triaxial shape for the remnant,
a much larger fraction of the total angular momentum
of the system could be radiated away through GW emission. In this case
the final merged core may resemble a Dedekind ellipsoid, i.e., it will have
a triaxial shape supported entirely by internal fluid motions, but with
a stationary shape in the inertial frame (so that it no longer
radiates GWs). The stability of such a configuration to gravitational
collapse has never been studied.
In contrast, in all 3-D numerical simulations performed to date,
the shear is quickly dissipated, so that GW emission
never gets a chance to extract more than a small fraction ($\sim10$\%)
of the angular momentum, and the final core appears to be 
nearly axisymmetric. However, this behavior could be an
artefact of the spurious shear viscosity present in all
3-D simulations. 

\section{Black Hole -- Neutron Star Mergers}

We now turn to BH--NS binaries, summarizing some recent (and
rather preliminary) results we have obtained with the Penn State
SPH code.
Fig.~4 shows why the final merger of a BH--NS binary
(for a typical stellar-mass BH) is particularly interesting
but also particularly difficult to compute: the tidal (Roche)
limit is typically right around the ISCO and the BH horizon.
On the one hand, this implies that careful, fully relativistic calculations
are needed. On the other hand, it also means that the fluid behavior
and the GW signals could
depend sensitively (and carry rich information) on both the masses and 
spins of the stars, and on the NS EOS. How much information is carried about
the fluid depends on where exactly the tidal disruption of the NS occurs:
for a sufficiently massive BH, the horizon will always be encountered
well outside the tidal limit, in which case the NS behavior remains
point-like throughout the merger.

% Fig 4
\begin{figure}[tbp]
\epsfxsize=10.5cm
\centerline{\epsfbox{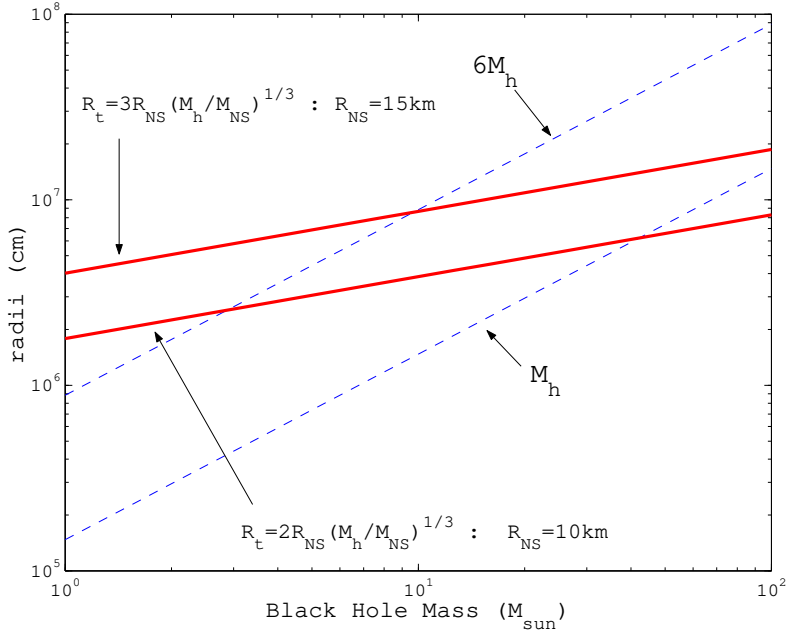}}
\caption{Tidal disruption limits for a $1.4\,M_\odot$ NS in circular orbit
around a BH of mass $M_{\rm h}$. The radius of the innermost
stable circular orbit (ISCO) is also shown for the Schwarzschild
and maximally corotating Kerr cases (dashed lines). The two 
solid lines bracket the Roche limit for NS with different spins
and radii.} 
\label{RL}
\end{figure}

We set up initial conditions for BH--NS binaries near the Roche limit
using the Penn State SPH code and a relaxation technique similar to
those used for previous SPH studies of close binaries (e.g.,
Rasio \& Shapiro 1995). First we construct hydrostatic equilibrium NS
models for a simple gamma-law EOS by solving the Lane-Emden
equation.  When the NS with this hydrostatic profile is placed in orbit
near a BH,
spurious motions could result as the fluid responds dynamically
to the sudden appearance of a strong tidal force. Instead, the initial
conditions for our dynamical calculations are obtained by relaxing the
NS in the presence of a BH in the corotating frame of the binary.    
For {\it synchronized\/} configurations (assumed here), the relaxation is done 
by adding an artificial
friction term to the Euler equation of
motion in the corotation frame. This forces the
system to relax to a minimum-energy state. We numerically 
determine the angular velocity $\Omega$ corresponding to a
circular orbit at a given $r$ as part of the relaxation process. 
The advantage of using SPH
itself for setting up equilibrium solution is that the dynamical
stability of these solutions can then be tested immediately by using
them as initial conditions for dynamical calculations.  

% Fig 5
\begin{figure}[tbp]
\epsfxsize=8.5cm
\centerline{\epsfbox{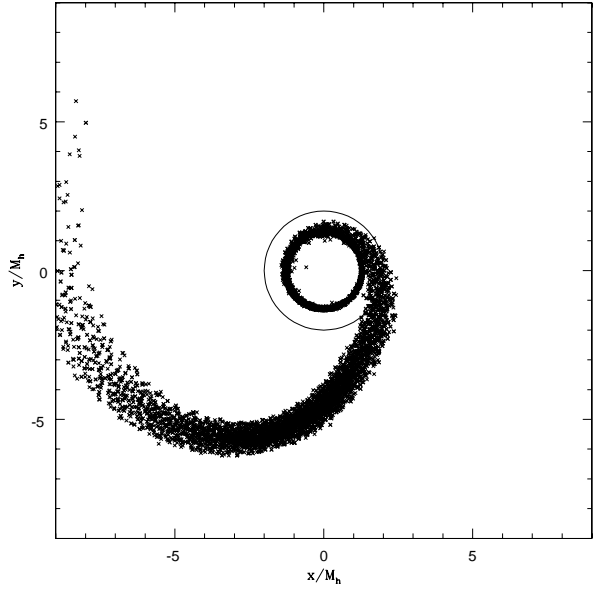}}
\caption{Tidal disruption of a $1.4\,M_\odot$ NS by a maximally corotating 
Kerr BH of mass $M_{\rm h}=10\,M_\odot$.
The NS was initially placed just outside the Roche limit ($r\simeq6.5\,M_{\rm h}$
for this NS of radius $R\simeq15\,$km with a $\Gamma=2$ polytropic EOS). In this snapshot near the end
of the simulation all SPH particles are shown projected onto the equatorial
plane of the system. The NS has been completely disrupted, and a ring
of material has formed just outside the BH horizon (at $r=M_{\rm h}$;
the circle drawn for reference corresponds to $r=2\,M_{\rm h}$).} 
\end{figure}

% Fig 6
\begin{figure}[tbp]
\epsfxsize=8.5cm
\centerline{\epsfbox{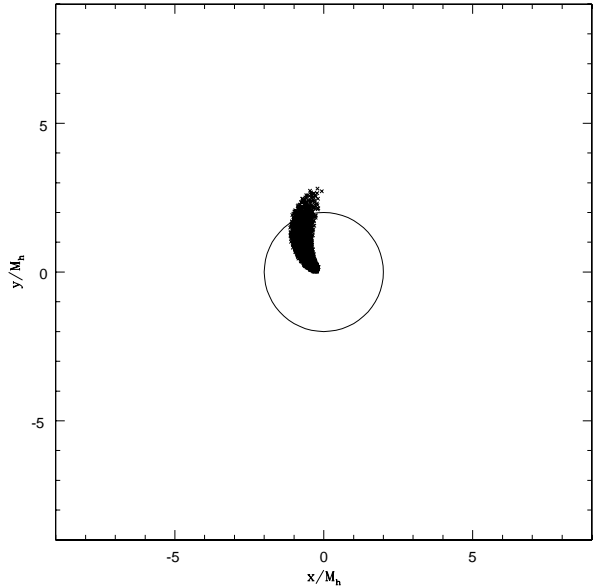}}
\caption{Same as in Fig.~5 but for a nonrotating (Schwarzschild) BH.
Here again the NS has been completely disrupted, but the NS fluid is
disappearing into the BH horizon (at $r=2\,M_{\rm h}$, indicated by circle)
and a stable ring of matter cannot form around the BH. Note that our
inner absorbing boundary is located at $r=0.2\,M_{\rm h}$, well inside the
horizon.} 
\end{figure}

We illustrate our results from two
simulations using $10^4$ SPH particles to represent a NS with a $\Gamma=2$ 
polytropic EOS. The only difference between these two simulations is in
the assumed spin of the BH: maximally corotating (Kerr parameter $a/M_{\rm h}=1$
and corotating NS in the equatorial plane) in one case; non-rotating (Schwarzschild BH) for the other. Snapshots of the SPH particles near the end of these 
simulations are shown in Figs.~5 and~6. Around the maximally
rotating BH, the fluid forms a stable ring just outside the BH horizon,
while for the non-spinning BH the entire mass of the NS disappears into the
horizon and is accreted by the BH on a dynamical timescale. There are
many obvious implications of these results for models of gamma-ray bursts,
which depend crucially on the presence of an accretion disk or
torus around the BH. We note also that, so far, we have never observed
the behavior seen in Newtonian simulations, where ``stable mass transfer''
from the NS onto the BH can persist for many orbits. Although our 
calculations so far
have been for initial circular orbits in the equatorial plane of a Kerr BH,
this is probably not realistic for most BH--NS binaries, where the NS is
formed second, with a kick that would have likely tilted the plane of the orbit
with respect to the BH spin (Kalogera 2000).

We have also performed several test calculations for white dwarfs (WDs) orbiting stably around a much more massive BH near their Roche limits. In this case the
mass ratio is extreme, and our approximation (moving the fluid on a fixed background metric) should be nearly exact. For example, if we place a $0.6\,M_\odot$ WD in orbit around
a Schwarzschild BH with $M_{\rm h}=2\times10^5\,M_\odot$, 
we find that we can maintain a circular
orbit at $r=8\,M_{\rm h}$ to within $|\Delta r|/r < 10^{-3}$ over one full 
orbital period. This is illustrated in Fig.~7.

% Fig 7
\begin{figure}[tbp]
\epsfxsize=8.5cm
\centerline{\epsfbox{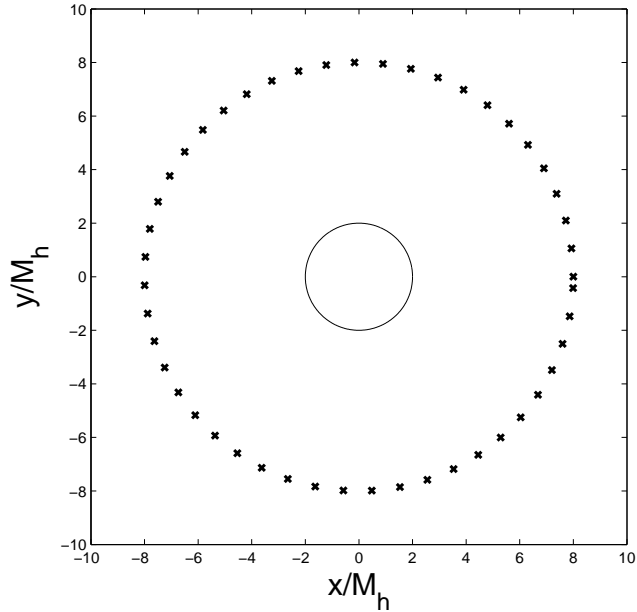}}
\caption{Test calculation for a WD orbiting a much more massive BH at 
$r=8\, M_{\rm h}$. The mass ratio in this case is $q\simeq 4\times10^5$.
The constant radius of this circular orbit is maintained 
by our code to within better than $10^{-3}$ over one full period.
Here each cross indicates the position of the WD (center of mass of
all SPH particles) at a 
different time along the orbit (counter-clockwise).} 
\end{figure}

\section*{Acknowledgements}

This work was supported by NSF Grant PHY-0245028 at Northwestern University.
SK and PL acknowledge support from the Eberly Research Funds of 
Pennsylvania State University, the Center for Gravitational Wave
Physics, funded under NSF cooperative agreement PHY-0114375, and a
NASA Swift Cycle~1 GI program. FR also thanks the Center for Gravitational Wave
Physics for hospitality and support. JF is supported by an NSF Postdoctoral
Fellowship in Astronomy and Astrophysics at UIUC.

\end{document}